\documentclass[traditabstract]{aa}  
\usepackage{geometry}
\usepackage{xcolor}
\usepackage{colortbl}
\usepackage{tabularx}
\usepackage{graphicx} 
\usepackage{multirow}
\definecolor{lightgray}{rgb}{0.83, 0.83, 0.83}
\definecolor{lightblue}{rgb}{0.67, 0.84, 0.90}
\definecolor{lightgreen}{rgb}{0.56, 0.93, 0.56}

\usepackage{natbib}
\bibpunct{(}{)}{;}{a}{}{,} 

\usepackage{graphics}   
\usepackage{graphicx}   
\usepackage{epstopdf}
\usepackage{longtable}   
\usepackage{url}      
\usepackage{bm}        
\usepackage{soul}

\usepackage[varg]{txfonts}
\usepackage{pdflscape}
\usepackage{supertabular}
\usepackage{hyperref}
\usepackage{xcolor}
\usepackage[normalem]{ulem}
\usepackage{natbib}
\usepackage{amsmath}
\definecolor{green}{rgb}{0.3,0.7,0.}
\definecolor{purple}{rgb}{0.77, 0.29, 0.55}
\newcommand\Mpy{\ensuremath{\ \msun\,{\rm yr}^{-1}}}

\newcommand\gva{{\sc Genec}}

\newcommand{\msolar} {$\rm{M_{\odot}}~$}
\newcommand{\msolarc} {$\rm{M_{\odot}}$}
\newcommand{\msolaryr} {$\rm{M_{\odot}~yr^{-1}}~$}
\newcommand{\msolaryrc} {$\rm{M_{\odot}~yr^{-1}}$}

\hypersetup{
    bookmarks=true,         
    unicode=false,          
    colorlinks=true,       
    linkcolor=blue,          
    citecolor=blue,        
    filecolor=blue,      
    urlcolor=blue           
}

\usepackage{color}

\newcommand{\msun}{\ensuremath{\mathrm{M}_{\odot}}\xspace}

\begin{document}

\title{Explaining the high nitrogen abundances observed in high-z galaxies via population~III stars of a few thousand solar masses}  

\titlerunning{CNO signatures \& PopIII Stars}
\author{Devesh Nandal\inst{1}, John A. Regan\inst{2}, Tyrone E. Woods\inst{3},  Eoin Farrell\inst{1}, Sylvia Ekström\inst{1}, Georges Meynet\inst{1}}
\authorrunning{Nandal et al.}

\institute{D\'epartement d'Astronomie, Universit\'e de Gen\`eve, Chemin Pegasi 51, CH-1290 Versoix, Switzerland \and
Centre for Astrophysics and Space Sciences Maynooth, Department of Theoretical Physics, Maynooth University, Maynooth, Ireland \and
Department of Physics and Astronomy, Allen Building, 30A Sifton Rd, University of Manitoba, Winnipeg MB  R3T 2N2, Canada}

\date{}

\abstract{ 

The chemical enrichment of the early Universe is a crucial element in the formation and evolution of galaxies, and Population III (Pop~III) stars must play a vital role in this process. In this study, we examine metal enrichment from massive stars in the early Universe's embryonic galaxies. Using radiation hydrodynamic simulations and stellar evolution modelling, we calculated the expected metal yield from these stars. Specifically, we applied accretion rates from a previous radiation-hydrodynamic simulation to inform our stellar evolution modelling, executed with the Geneva code, across 11 selected datasets, with final stellar masses between 500 and 9000 \(\mathrm{M}_{\odot}\).
Our results demonstrate that the first generation of Pop III stars within a mass range of 2000  to \(9000 \, \mathrm{M}_{\odot}\) result in N/O, C/O and O/H ratios compatible with the values observed in 
very high-z galaxies GN-z11 and CEERS 1019. The ejecta of these Pop III stars are predominantly composed of \(^4\)He, \(^1\)H, and \(^{14}\)N. Our Pop III chemical enrichment model of the halo can accurately reproduce the observed N/O and C/O ratios, and, by incorporating a hundred times more zero-metallicity interstellar material with the stellar ejecta, it accurately attains the observed O/H ratio. Thus, a sub-population of extremely massive Pop~III stars, with masses surpassing approximately 2000 M$_{\odot}$, effectively reproduces the CNO elemental abundances observed in high-z JWST galaxies to date.
We closely reproduced the observed Ne/O ratio in CEERS 1019 employing a model with several thousand solar masses and non-zero metallicity, and we projected a \(^{12}C/^{13}C\) ratio of 7, substantially lower than the solar ratio of around 90.
The significant nitrogen enrichment predicted by Pop III stars with a few thousand solar masses not only reinforces the argument for a heavy seed formation pathway for massive black holes at redshifts as high as z=10.6 but it also accentuates the need for deeper investigations into their complex nature and pivotal role in the early Universe.
}


\maketitle

\section{Introduction}

The observation of nitrogen-enhanced high-$z$ galaxies, particularly in GN-z11 \citep{Oesch2016} and CEERS-1019 \citep{Finkelstein2017}, has sparked a revolution in the study of the Universe's earliest star and galaxy formation. GN-z11, located at a redshift of $\mathrm{z} = 10.6$ \citep{Bunker2023}, exhibits a N/O abundance ratio in its interstellar medium (ISM) that is more than four times the solar value ($\log(\mathrm{N/O}) >\sim -0.25$) \citep{Cameron2023, Senchyna2023}. This ratio is not only significantly higher than what is typically found in low-redshift galaxies and HII regions with a comparable metallicity ($12+\log({\rm O/H}) <\sim 8.0$) but it also slightly exceeds the values observed in galaxies with super-solar metallicities \citep{Vincenzo2016, Berg2019}. Although the C/O ratio of GN-z11 is consistent with normal values, it is poorly constrained \citep{Cameron2023}. 

Similarly, CEERS-1019, at a redshift of $z=8.678$, displays strong nitrogen emission lines of N III and N IV, with the N IV] $\lambda1486$ being the most intense line in its rest-frame UV spectrum \citep{Tang2023}. This galaxy has an unusually high N/O abundance ratio of $\log(\mathrm{N/O})= -0.13 \pm 0.11$, which is approximately 5.6 times the solar ratio. However, its C/O and Ne/O ratios, with values of \(\log(\mathrm{C/O})= -0.75 \pm 0.11\) and \(\log(\mathrm{Ne/O})= -0.63 \pm 0.07\)
 respectively, are relatively normal for its metallicity \citep{Marques2023}.

These observations challenge the conventional theories of stellar yields and galactic chemical evolution, necessitating a reexamination of the dynamics of the early Universe. In particular these unusual abundance patterns in high-z galaxies may be a signature of  massive, very massive, and supermassive stars (MSs, VMSs, and SMSs) \footnote{We consider stars to be massive within  10-100 M$_\odot$, very massive within 100-1000 M$_\odot$, extremely massive within 1000-40,000 M$_\odot$ and supermassive with mass above 40,000 M$_\odot$ as described in the work by \citet{Woods2020}} in early galaxies. Moreover, these massive objects may well explain the appearance of central massive black holes in these galaxies as well \citep[e.g.][]{Natarajan_2023} and their role in the formation of supermassive black holes (SMBHs) in the early Universe is a subject of ongoing research \citep{Hosokawa2010, Ekstrom2012, Woods2017, Lionel2018, Woods2021}. \\

\indent A number of recent studies have attempted to explain the unusually high Nitrogen abundances (as well as the other metal abundances) observed in these high-z galaxies. 
\citet{Kobayashi2023} explain these signatures using a rapid chemical enrichment mechanism via a dual-burst model where a quiescent period lasting approximately 100 Myr separates the bursts. In their model, Wolf-Rayet stars become the dominant enrichment source during the second burst  - enriching the ISM with heavy elements. Their single burst model, similar to ours described here, requires the presence of massive stars with masses in excess of $\sim 1000$ M$_\odot$. Along similar lines both \citet{Charbonnel2023} and \citet{Nagele2023} have studied the impact that metal enriched stars in the mass range of $10^3$ - $10^5$ M$_\odot$ can have in high-z galaxies. In both of these studies the authors have been able to reproduce the high $\log(\mathrm{N/O})$ abundance ratios observed in GN-z11. Several other studies have sought to explore different stellar evolution channels on the early chemical evolution of the Universe. For instance, \citet{Meynet2006} and \citet{Choplin2018} investigated the impact of fast-rotation on the nucleosynthetic outputs of  massive stars. These yields were then used in models for the early chemical evolution of the Milky Way by \citet{Chiap2003, Chia2006, Chia2013} showing that the chemical composition of the bulk of halo stars is compatible with yields coming from fast rotating models.
The potential role of pair-instability supernovae in the early Universe was analysed by \citet{Heger2002}. Although these studies have shed light on different aspects, the exact elemental yields and enrichment potential of VMSs and SMSs remain largely unexplored \citep{Denissenkov2014}.  

In this paper, we adopt a novel approach by combining radiation hydrodynamic simulations and stellar evolution modelling to compute the expected metal yield from a population of massive Population III (PopIII) stars. Using the accretion rates from \citet{Regan_2020b}, we compute eleven PopIII models from the pre-main sequence (PreMS) until the end of core helium burning. We delve into the results of individual stellar models and the integrated halo population, revealing new insights into the N/O, C/O, and O/H number fractions, and taking into account the possibly diluting impact of an additional population of one hundred 20 \(\mathrm{M}_{\odot}\) stars. Specifically, we find that chemical enrichment coming from the ejecta of very massive stars, computed with a variety of assumptions can reproduce the observed N/O number fractions, while we successfully match the observed C/O abundances when pulsations from the massive PopIII stars remove all mass above the CO core. Finally to match the O/H number fractions we require a dilution factor (given by the ratio of the mass of the ejecta to that of the interstellar medium with which the ejecta mixed) of 100, entirely consistent with expected ratios chemical abundances in gas rich high-z galaxies. This study not only provides significant insights into the chemical enrichment of the early Universe and the role of supermassive PopIII stars, but also underscores the necessity for further investigations into the complex but pivotal role of massive PopIII stars in the early Universe.

This paper is set out as follows: Section \ref{Sec:Results} first explores the mechanisms invoked for the transport of chemical special and calculates the integrated abundances of the individual stars (we express $\mathbf X_{\mathbf {io}}$ as the mass fraction of element $i$ in the ejecta a given star) and for the entire halo ($\mathbf X_{\mathbf {ip}}$, the mean mass fraction of element $i$ in the ejecta coming from a population of stars). Section \ref{Sec:Discussion} explores the mass ranges of individual stars that would reproduce the observed results and predicts abundances based on our models, along with comparing our results to previous studies. Section \ref{Sec:Conclusion} sums up our findings and provides prospects for future work.

\section{Methods} \label{Sec:Methods}
\subsection{Cosmological context} \label{Sec:Context}
As discussed in the Introduction, the goal of this study is to calculate the metal enrichement due to pulsational mass loss from massive stars in embryonic haloes in the early Universe. To achieve this goal we utilise the radiation hydrodynamical simulation of \citet{Regan_2020b}. We previously used the calculated accretion rates from \citet{Regan_2020b} to calculate the critical accretion rates onto massive PopIII stars in the study by \citet{Nandal2023}. In this study we take the stars found in \citet{Regan_2020b} and compute, in post-processing, the expected metal yield from these stars using the same stellar evolution code as used in \citet{Nandal2023}.\\
\indent For an in-depth discussion of both \citet{Regan_2020b} and \citet{Nandal2023} we refer readers to the original papers but describe both papers briefly here to provide context. \citet{Regan_2020b} employed cosmological zoom-in simulations utilising the \texttt{Enzo} code \citep{Enzo_2014, Enzo_2019} to investigate pristine, atomic cooling haloes, from the original studies by \cite{Wise_2019} and \cite{Regan_2020}\footnote{These simulation were themselves rooted in the \textit{Renaissance} simulation suite \citep{Chen_2014, OShea_2015, Xu_2016b}}. Using the canonical zoom-in technique \citet{Regan_2020b} study showed the formation of 101 (massive) stars in one single halo. The cumulative stellar mass reached roughly 90,000 \msolar after about 2 Myr of the first star's inception. Stellar masses ranged between 40 \msolar to over 6000 \msolarc, with a simulation spatial resolution of $\Delta x \sim 1000$ au. Accretion rates onto individual stars ranged from less than $10^{-3}$ \msolaryr to close to 1 \msolaryr at peak. 
It is these accretion rates which serve as the input for our stellar evolution modelling and is a key and unique component in this study.

In \citet{Nandal2023} we selected 11 stellar evolution datasets from the underlying cosmological simulations of R20b comprised of a variety of accretion histories and final stellar masses (see Table \ref{table:11models}). We use these same 11 datasets here. They contain final stellar masses from 491 \msolarc \ up to 6127 \msolarc. These 11 stellar models, with fluctuating accretion rates, were evolved from the pre-MS to the end of core helium burning using the Geneva code (\gva) \citep{Eggenberger2008}. We adopted a consistent chemical mix ($X$ = 0.7516, $Y$ = 0.2484, $Z$ = 0) and a deuterium mass fraction $X_{2} = 5 \times 10^{-5}$ as per previous studies. Starting with a 2 M$_{\odot}$ fully convective seed, accretion is modeled via a thin cold disc \citep{Lionel2013, Lionel2016}, consistent with \cite{Palla1992} \& \cite{Hosokawa2010}. The original study by \citet{Nandal2023} required several code enhancements in \gva~ including functionality to cater for variable accretion rates from external files as well as improved spatial and time resolution to account for the variable accretion history. However, effects due to rotation and (explicit) mass loss are not considered, although we note that our model hinges on the assumption that these massive stars do suffer from some degree of mass loss through for example pulsations or due to stellar winds triggered by the increase of the opacity at the surface due to changes of the surface composition here by convection (see Figure \ref{Fig:Illu} for a graphical illustration). In this work, we do not model the mass loss numerically, but test various hypothesis and assume that some amount occurs during the evolution. 

We discuss these underlying assumptions further in \S \ref{Sec:classic}. In addition to the 11 models already computed in \citet{Nandal2023} we computed an additional model with a constant accretion rate of 3.5$\times 10^{-3}$\Mpy and a final mass of 8904 M$_{\odot}$ using the same chemical composition as above. This model was computed to serve as a limiting case for halo from the simulations of \citet{Regan_2020b} to match the N/O, C/O and O/H values with GN-z11 and CEERS 1019.  

\subsection{Massive PopIII Star (Nitrogen) Enrichment}\label{Sec:classic}

\begin{figure*}[!h]
	\centering
		\includegraphics[width=18cm]{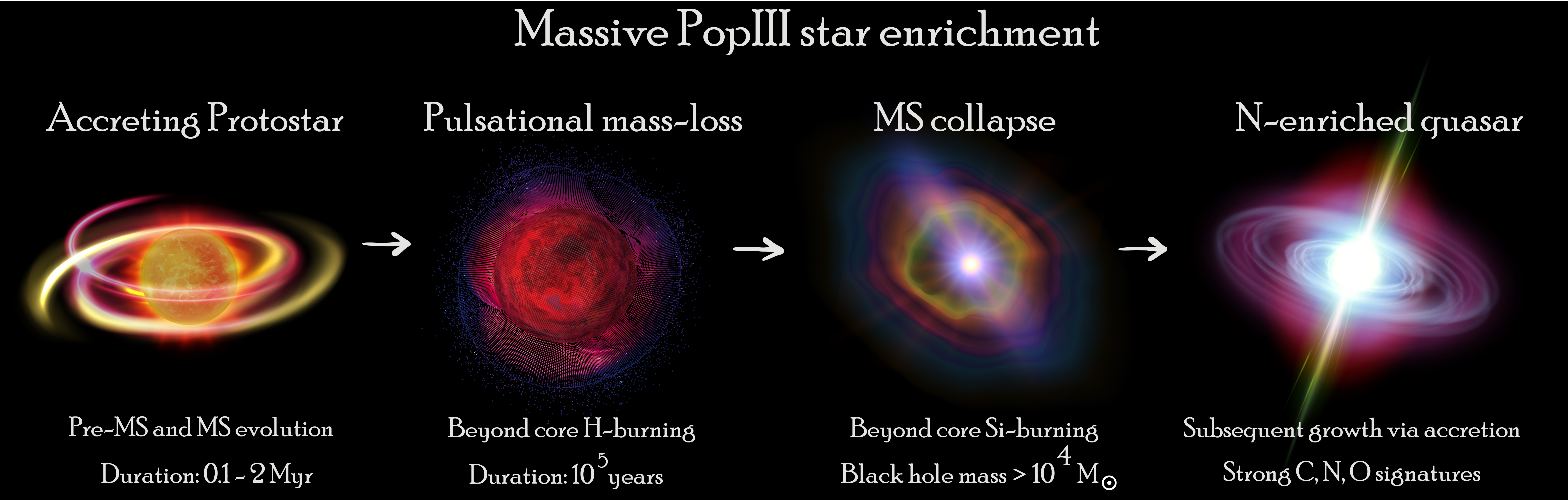}
	\caption{Illustration depicting the massive PopIII enrichment scenario. From left to right: (1) The formation of a massive PopIII star via accretion which generally occurs during the pre-MS but some stars may undergo short accretion phases during MS. If the accretion rate is below the critical limit described in \citet{Nandal2023}, the star migrates to blue. (2) The transition to red at the end of core helium burning leads to the formation of large convective zones and pulsational mass loss as shown by the red mesh. (3) The collapse of a MS may lead to additional mass ejected and the eventual formation of a black hole. (4) Further accretion of ejected material onto the black hole creats the conditions for allowing nitrogen absorption lines to appear, as in the case of GN-z11 and CEERS 1019.
 }
		\label{Fig:Illu}
\end{figure*}

To explain the abundances observed in high redshift candidates such as GN-z11 and CEERS1019, we invoke the very and extremely massive PopIII enrichment scenario. In this scenario, a very and extremely massive star begins its evolution as an accreting protostar during the pre-Main Sequence (pre-MS) evolution that lasts for $\lesssim 10^5$ yr, see Figure \ref{Fig:Illu}. As the star transitions to the main sequence the core hydrogen burning phase begins in the blue with the star possessing a convective core and a radiative envelope as shown in detail in \citet{Nandal2023}. If the star accretes sufficiently rapidly (above approximately $2.5 \times 10^{-3}$ \msolaryrc) then the star can transition to the red during the pre-MS or the MS phase. This transition is included in our modelling. In either case at the end of the MS the star migrates to the red and acquires a near fully convective structure before the start of core helium burning. This allows the efficient transport of chemical species from the interior to the surface. We next make the assumption, that due to the unstable nature of such massive PopIII stars, they are unstable to mass loss, as suggested by \citet{Nagele2022}. The quantity of mass lost via winds and/or pulsations is difficult to quantify and instead we parameterise this unknown quantity by examining the impact of different mass loss amounts (see \S \ref{Sec:Abund})\footnote{A similar approach has been used by \citet{Liu2021} for investigating the impact of winds of Pop III massive stars.}. 
Following Helium burning and at the end of core silicon burning, such an object may explode as a supernova \citep{Nagele2023} and produce a black hole  or directly collapse into a black hole with mass comparable to the Helium core mass \citet{Montero_2012}. For massive stars with masses in excess of approximately 260 \msolar a direct collapse black hole is always the expected outcome. However, small mass islands leading to a supernova explosions are predicted in some cases \citep[e.g.][]{Chen_2014, Moriya_2021} but there is currently no observational guidance to constrain these islands. For the case of massive PopIII stars with masses in the range of approximately $10^3 - 10^4$ \msolar we find that the expulsion of large quantities of surface material, which contains unusually large nitrogen fractions, from the surface can lead to a black hole accretion disc with an unusually large nitrogen abundance. This key signature of massive PopIII mass loss episodes is highly relevant to other recent observations of N-rich quasars as described in the works of \citep{Bunker2023,Cameron2023}. 
\subsection{Model for the wind enrichment}\label{Sec:ratio}
In this study, we study a scenario where a star enriches the chemical composition of its surrounding interstellar medium (ISM) by shedding some of its mass during its evolution. This mass loss might occur through a series of significant mass-loss episodes prior to the star's final collapse for example. Our understanding of the mass loss rates of present day massive stars is incomplete \citep{Sabhahit2023} and in the absence of observational guidance our understanding of early Universe massive stars should certainly be seen as incomplete. We therefore choose to parameterise the mass-loss rates of these stars. We detail our parameterisation as follows.\\

\indent We assume that the stars lose their mass at or after the core He-burning phase. This is a phase where the star is in the red part of the HR diagram where pulsational instabilities can be large and also a phase where the surface is enriched in heavy elements increasing the opacity.
The total mass of element $\mathbf{i}$ ejected by the star, denoted as $\mathbf{{M_{EJ}}(i)}$ in solar masses, is given by $\rm{\bar{X}_{i} M_{\rm EJ}}$, where $\rm{\bar{X}{i}}$ is the mean mass fraction of element $\mathbf{i}$ in the ejecta of the star and M$_{\rm {EJ}}$ is the total mass ejected. $\rm{\bar{X}_{i}}$ is given by 
\begin{equation}
\rm{\bar{X}_{i}}=\int_{M_{\rm cut}}^{M_{\rm tot}} X_{i}(M_{r}) , dM_r/M_{\rm EJ}
\label{eq:1}
\end{equation}
where $\mathbf{M_{\mathbf cut}}$ is the mass coordinate above which we assume the mass to be lost, $\mathbf M_\mathbf{r}$ is the mass coordinate at which $\mathbf X{\mathbf i}$ is chosen. Here we assume $\mathbf{M_{\mathbf cut}}$ to take on three separate values: 1) 10\% of the total stellar mass; 2) 40\% of the total stellar mass; and 3) all the stellar mass above CO core. The CO core is defined as the mass coordinate below which less than $10^{-3}$ of Helium mass fraction ($\mathbf Y_c$) is present. The mass below $\mathbf{M_{\mathbf cut}}$ is therefore assumed to be locked into the black hole, and $\mathbf{M_{\mathbf tot}}$ is the total actual mass of the star at the moment of core collapse.

 The ejected mass interacts with the interstellar medium (ISM), leading to a mixture of the star's ejecta and the ISM. The mass of the ISM involved in this interaction is represented as $\mathbf{M_{\mathbf ISM}}$. We introduce a dilution factor $\mathbf{f}$, which is the ratio $\rm{M_{ISM}/M_{EJ}}$. As noted we treat $\mathbf{f}$ as a free parameter as it depends on a number of unknown factors, such as the physics of the explosion and/or the collapse on the one hand and on the exact state of the ISM on the other. To account for our ignorance of both the exact mass lost values and the chemical composition of the ISM we explore different possibilities here considering this $\mathbf{f}$ factor as a free parameter, varying from 1 to 1000. We determine $\mathbf X_{\mathbf {io}}$ as the mass fraction of element $\mathbf{i}$ in the ISM at ejection time. The ejection of the mass from a star is assumed to occur in the early stages of the halo evolution. As stars continue to eject matter in their surroundings, the chemical composition changes as a function of time. In other words, the initial chemical composition of the PopIII star is pristine and only consists of hydrogen and helium but as these stars evolve and eject the CNO rich material in the ISM, the chemical composition of the ISM in the halo evolves as a function of time. 

Using these parameters, we estimate the final composition of the chemical mixture of the star's ejecta and the ISM. This is a key objective of our study. We define $\mathbf X_{\mathbf {im}}$ as the mass fraction of element $\mathbf{i}$ in this mixture and calculate it using the principle of mass conservation:

\begin{equation}
\mathrm{X_{im} * (M_{EJ} + M_{ISM}) = \bar{X}_{i}* M_{EJ} + X_{io}* M_{ISM}}
\end{equation}

\noindent By substituting the dilution factor f = M$_{\rm ISM}$/M$_{\rm EJ}$ into the equation, we can simplify it to:

\begin{equation}
\mathrm{X_{im}*(1+f) = \bar{X}_{i} + f*X_{io}}
\end{equation}

Furthermore, to extrapolate this to a galaxy containing $n$ stars is straight forward. Let $\mathbf {n ({\rm M}_{\rm ini})}$ be the number of stars in the galaxy with masses between M$_{\rm ini}$ and M$_{\rm ini}$ + dM$_{\rm ini}$. We obtain the mean mass fraction of element $\mathbf{i}$ in the ejecta of all stars in the galaxy as

\begin{equation}
\mathrm{X_{ip} = \sum_{M_{ini}} \frac{n(M_{ini}) * \bar{X}_{i}(M_{ini}) * M_{EJ}(M_{ini})}{n(M_{ini}) * M_{EJ}(M_{ini})}}
\label{eq:4}
\end{equation}

where $\mathbf X_{\mathbf {ip}}$ is the mass fraction of element $\mathbf{i}$ in the ejecta of all the stars. 
Now again assuming the dilution $\mathbf f$ as before we can compute the mass fraction in the ISM following the enrichment from the stars as 
\begin{equation}
\mathrm{X_{imp} = \frac{X_{ip} + f*X_{io}}{1 + f}}
\end{equation}

Finally, we assume the formation of stars in the galaxy occurs at zero metallicity and therefore $\mathbf X_{\mathbf {io}}$ is taken as zero for C, N, O and 0.7516 for $\mathbf X_{\mathbf {H, ini}}$. 
This means that for all elements excluding Hydrogen we have that
\begin{equation}
\mathrm{X_{imp} \approx \frac{X_{ip}}{1+f}}
\end{equation}

Therefore, the mass fraction of element $\mathbf i$ that exists in the ISM following enrichment
is given by the ejected fraction, $\mathbf X_{\mathbf {ip}}$, divided by the dilution factor, $\mathbf f$, if  $\mathbf f$ is much larger than 1. \textbf{The initial mass function for the halo is given by the simulation, that is, the histograms shown in Fig. \ref{Fig:M_ejecta}}

The goal of this paper is to systematically explore the various integrated abundances (i.e. $\mathbf X_{\mathbf {imp}}$) expected from very massive PopIII stars in the first galaxies.

\begin{table*}[t]
\caption{Mean abundances in the mass fraction of different chemical species in the mass ejected by 11 massive PopIII star models. The first column shows the final mass of each model and the second column depicts the mass of CO core at the end of core helium burning. The next 5 columns depict the mean abundances (see Eq.\ref{eq:1} in the mass fraction for $^{1}H$, $^{4}He$, $^{12}C$, $^{14}N$ and $^{16}O$ when the outermost 10\% of total mass is ejected. The following five columns are the mean abundance in the ejecta when 40\% of the outermost total mass is ejected and the final five columns when all the mass above CO core is ejected.} 
\centering
\begin{tabularx}{\textwidth}{*{17}{X}}
\\ \hline
 & & & & & & & & & & & & & & & & \\
$M_{\text{tot}}$  & $M_{\text{CO}}$  & $\bar{X}_{\text{H}}$ & $\bar{X}_{\text{He}}$ & $\bar{X}_{\text{C}}$ & $\bar{X}_{\text{N}}$ & $\bar{X}_{\text{O}}$ & $\bar{X}_{\text{H}}$ & $\bar{X}_{\text{He}}$ & $\bar{X}_{\text{C}}$ & $\bar{X}_{\text{N}}$ & $\bar{X}_{\text{O}}$ & $\bar{X}_{\text{H}}$ & $\bar{X}_{\text{He}}$ & $\bar{X}_{\text{C}}$ & $\bar{X}_{\text{N}}$ & $\bar{X}_{\text{O}}$ \\ \hline
$(M_\odot)$ & $(M_\odot)$ & \multicolumn{5}{c}{\cellcolor{lightgray}10\%} & \multicolumn{5}{c}{\cellcolor{lightblue}40\%} & \multicolumn{5}{c}{\cellcolor{lightgreen}CO core} \\ \hline
491 & 104 & \cellcolor{lightgray}0.360 & \cellcolor{lightgray}0.430 & \cellcolor{lightgray}0.002 & \cellcolor{lightgray}0.056 & \cellcolor{lightgray}0.150 & \cellcolor{lightblue}0.256 & \cellcolor{lightblue}0.450 & \cellcolor{lightblue}0.002 & \cellcolor{lightblue}0.079 & \cellcolor{lightblue}0.210 & \cellcolor{lightgreen}0.235 & \cellcolor{lightgreen}0.451 & \cellcolor{lightgreen}0.002 & \cellcolor{lightgreen}0.084 & \cellcolor{lightgreen}0.224 \\
771 & 165 & \cellcolor{lightgray}0.319 & \cellcolor{lightgray}0.460 & \cellcolor{lightgray}0.002 & \cellcolor{lightgray}0.053 & \cellcolor{lightgray}0.162 & \cellcolor{lightblue}0.235 & \cellcolor{lightblue}0.466 & \cellcolor{lightblue}0.002 & \cellcolor{lightblue}0.074 & \cellcolor{lightblue}0.218 & \cellcolor{lightgreen}0.217 & \cellcolor{lightgreen}0.465 & \cellcolor{lightgreen}0.002 & \cellcolor{lightgreen}0.079 & \cellcolor{lightgreen}0.232 \\
778 & 163 & \cellcolor{lightgray}0.299 & \cellcolor{lightgray}0.473 & \cellcolor{lightgray}0.003 & \cellcolor{lightgray}0.038 & \cellcolor{lightgray}0.180 & \cellcolor{lightblue}0.240 & \cellcolor{lightblue}0.454 & \cellcolor{lightblue}0.002 & \cellcolor{lightblue}0.058 & \cellcolor{lightblue}0.235 & \cellcolor{lightgreen}0.220 & \cellcolor{lightgreen}0.447 & \cellcolor{lightgreen}0.002 & \cellcolor{lightgreen}0.065 & \cellcolor{lightgreen}0.253 \\
932 & 247 & \cellcolor{lightgray}0.339 & \cellcolor{lightgray}0.588 & \cellcolor{lightgray}0.002 & \cellcolor{lightgray}0.056 & \cellcolor{lightgray}0.015 & \cellcolor{lightblue}0.185 & \cellcolor{lightblue}0.682 & \cellcolor{lightblue}0.002 & \cellcolor{lightblue}0.102 & \cellcolor{lightblue}0.028 & \cellcolor{lightgreen}0.161 & \cellcolor{lightgreen}0.695 & \cellcolor{lightgreen}0.003 & \cellcolor{lightgreen}0.108 & \cellcolor{lightgreen}0.031 \\
1135 & 440 & \cellcolor{lightgray}0.204 & \cellcolor{lightgray}0.669 & \cellcolor{lightgray}0.003 & \cellcolor{lightgray}0.086 & \cellcolor{lightgray}0.038 & \cellcolor{lightblue}0.124 & \cellcolor{lightblue}0.691 & \cellcolor{lightblue}0.003 & \cellcolor{lightblue}0.122 & \cellcolor{lightblue}0.058 & \cellcolor{lightgreen}0.115 & \cellcolor{lightgreen}0.691 & \cellcolor{lightgreen}0.005 & \cellcolor{lightgreen}0.126 & \cellcolor{lightgreen}0.063 \\
1331 & 317 & \cellcolor{lightgray}0.293 & \cellcolor{lightgray}0.519 & \cellcolor{lightgray}0.005 & \cellcolor{lightgray}0.067 & \cellcolor{lightgray}0.114 & \cellcolor{lightblue}0.201 & \cellcolor{lightblue}0.528 & \cellcolor{lightblue}0.003 & \cellcolor{lightblue}0.103 & \cellcolor{lightblue}0.163 & \cellcolor{lightgreen}0.186 & \cellcolor{lightgreen}0.528 & \cellcolor{lightgreen}0.003 & \cellcolor{lightgreen}0.108 & \cellcolor{lightgreen}0.172 \\
1662 & 630 & \cellcolor{lightgray}0.198 & \cellcolor{lightgray}0.766 & \cellcolor{lightgray}0.001 & \cellcolor{lightgray}0.031 & \cellcolor{lightgray}0.004 & \cellcolor{lightblue}0.095 & \cellcolor{lightblue}0.852 & \cellcolor{lightblue}0.001 & \cellcolor{lightblue}0.047 & \cellcolor{lightblue}0.004 & \cellcolor{lightgreen}0.078 & \cellcolor{lightgreen}0.864 & \cellcolor{lightgreen}0.002 & \cellcolor{lightgreen}0.050 & \cellcolor{lightgreen}0.006 \\
1923 & 746 & \cellcolor{lightgray}0.232 & \cellcolor{lightgray}0.577 & \cellcolor{lightgray}0.001 & \cellcolor{lightgray}0.046 & \cellcolor{lightgray}0.137 & \cellcolor{lightblue}0.181 & \cellcolor{lightblue}0.573 & \cellcolor{lightblue}0.001 & \cellcolor{lightblue}0.061 & \cellcolor{lightblue}0.174 & \cellcolor{lightgreen}0.171 & \cellcolor{lightgreen}0.571 & \cellcolor{lightgreen}0.001 & \cellcolor{lightgreen}0.064 & \cellcolor{lightgreen}0.183 \\
3053 & 650 & \cellcolor{lightgray}0.202 & \cellcolor{lightgray}0.737 & \cellcolor{lightgray}0.000 & \cellcolor{lightgray}0.018 & \cellcolor{lightgray}0.004 & \cellcolor{lightblue}0.202 & \cellcolor{lightblue}0.737 & \cellcolor{lightblue}0.000 & \cellcolor{lightblue}0.022 & \cellcolor{lightblue}0.004 & \cellcolor{lightgreen}0.157 & \cellcolor{lightgreen}0.765 & \cellcolor{lightgreen}0.001 & \cellcolor{lightgreen}0.033 & \cellcolor{lightgreen}0.012 \\
4477 & 759 & \cellcolor{lightgray}0.297 & \cellcolor{lightgray}0.687 & \cellcolor{lightgray}0.000 & \cellcolor{lightgray}0.015 & \cellcolor{lightgray}0.001 & \cellcolor{lightblue}0.153 & \cellcolor{lightblue}0.816 & \cellcolor{lightblue}0.001 & \cellcolor{lightblue}0.029 & \cellcolor{lightblue}0.001 & \cellcolor{lightgreen}0.073 & \cellcolor{lightgreen}0.884 & \cellcolor{lightgreen}0.003 & \cellcolor{lightgreen}0.034 & \cellcolor{lightgreen}0.006 \\
6127 & 1252 & \cellcolor{lightgray}0.173 & \cellcolor{lightgray}0.779 & \cellcolor{lightgray}0.001 & \cellcolor{lightgray}0.039 & \cellcolor{lightgray}0.006 & \cellcolor{lightblue}0.079 & \cellcolor{lightblue}0.855 & \cellcolor{lightblue}0.001 & \cellcolor{lightblue}0.056 & \cellcolor{lightblue}0.008 & \cellcolor{lightgreen}0.040 & \cellcolor{lightgreen}0.875 & \cellcolor{lightgreen}0.004 & \cellcolor{lightgreen}0.060 & \cellcolor{lightgreen}0.017 \\
\hline
\end{tabularx}
\label{table:11models}
\end{table*}

\begin{figure*}[!t]
	\centering
		\includegraphics[width=8.7cm]{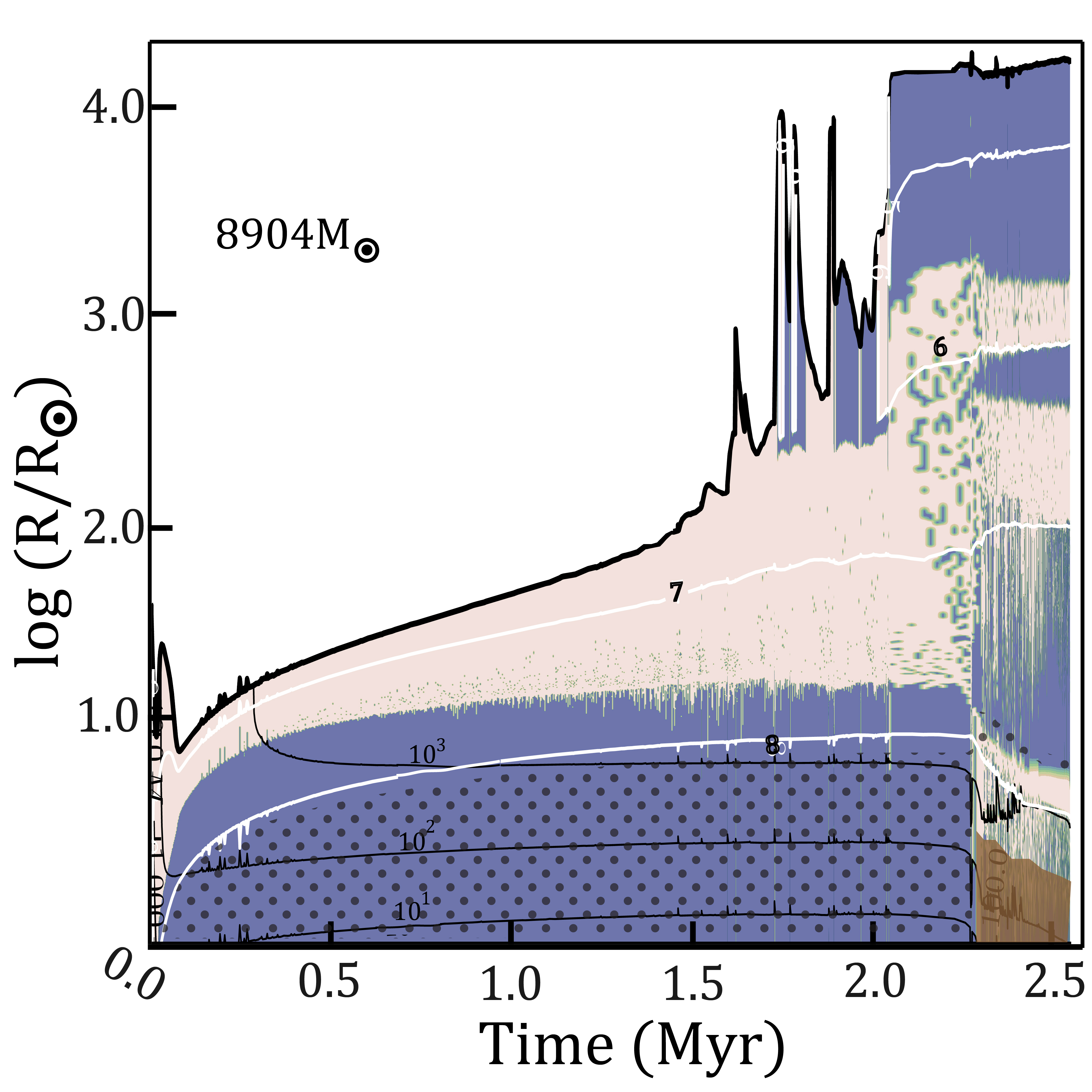}\includegraphics[width=9cm]{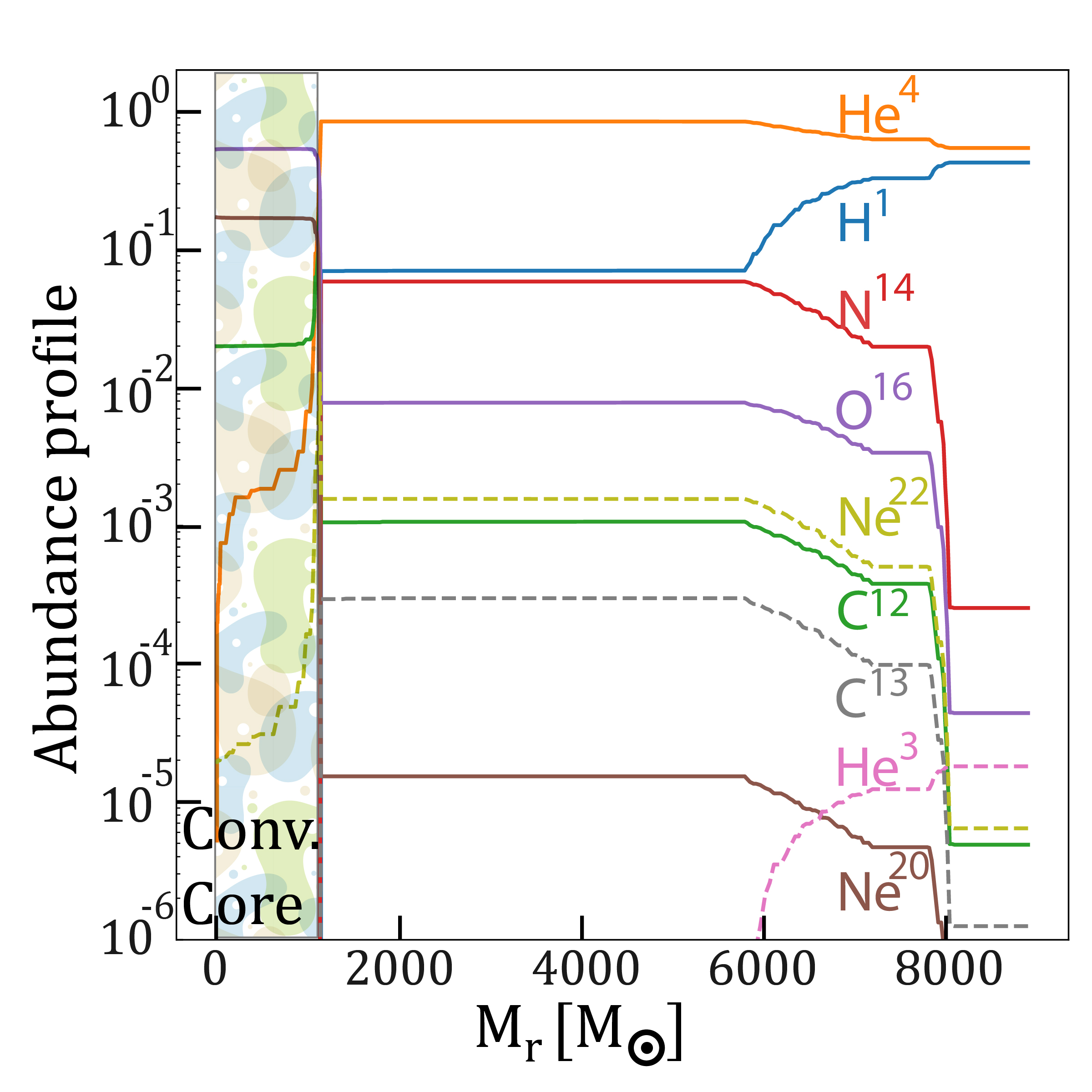}
	\caption{ Test model with a final mass of 8904 M$_{\odot}$ computed using a constant accretion rate of 3.5\Mpy until the end of core helium burning. {\it Left panel:} The Kippenhahn diagram showing the evolution of radius versus time. The blue and cream regions represents the convective and radiative zones respectively. Core hydrogen burning is shown in dotted zone and core helium burning in brown zone. The black lines represents the iso-masses and the white lines depict the iso-thermal lines. {\it Right panel:} The abundance profile inside the 8904 M$_{\odot}$ model at the end of core helium burning. The label for each element matches the corresponding line.
 }
		\label{Fig:Kipp}
\end{figure*}

\section{Results}\label{Sec:Results}

\subsection{Transport of chemical species}\label{Sec:Transport}
The evolution of a typical accreting massive Pop III star is shown in left panel of Figure \ref{Fig:Kipp}. The accretion history of this model is artificially generated by assuming a constant accretion rate of 3.5$\times 10^{-3}$\Mpy to produce a star with mass of 8904 M$_\odot$. The choice of constant accretion rate is motivated by the works of \citet{Nandal2023b} where they compare the effects of a constant versus variable accretion rate and find that evolution beyond core hydrogen burning is near identical in both cases. Therefore, we present a new model (in addition to the 11 models with variable accretion rates found in \citet{Nandal2023}) that reproduces the same chemical structure in advanced evolutionary stages as the model computed with a variable accretion rate ending with a mass of 8904 M$_\odot$. \textbf{It must be noted that the accretion history of this model was not obtained from the hydrodynamic simulations and therefore, the contribution of its stellar yields to the entire sub-population of mini halo are not included in the ejecta calculations. We simply use this model to explain the structure of such stars and explore its individual N/O, C/O and O/H ratios later on in this study}.
After following a pre-MS evolution of $\lesssim 10^5$ years, the object commences hydrogen burning in the blue region of HR diagram at log (T$_{\rm eff}) = 4.90$
\citep{Nandal2023}. During the middle of hydrogen burning, at an age of about 1.5 Myr the star begins its redward migration. The accretion rate during this migration is below the critical value of $\dot{M}_{\mathrm{crit, MS}}$ = 7.0$\times 10^{-3}$ found by \citet{Nandal2023}, and the star therefore oscillates between red and blue until $\approx$ 2 Myrs. As the core hydrogen is depleted, the star migrates to the red and attains an effective temperature of log (T$_{\rm eff}) = 3.70$. During this final crossing of the HR gap, the structure that previously consisted of a convective core and an extended radiative envelope, now consists of a convective core, a series of intermediate convective zones and an extended convective envelope. Such a near-fully convective structure allows the transport of chemical species from the deep interior to the surface, thereby promoting strong enrichment without the need of rotation. It is interesting to note that the near-fully convective structure is only attained at the very end of core hydrogen burning and is absent during the red-blue excursions that occur from 1.5 - 2 Myrs, see left panel of Figure \ref{Fig:Kipp}. The non-rotating models produced in this study using \gva \ show that the structure of such stars is independent of the choice of mass within the mass range of 200 - 10$^5$ \msun.

The right panel of Figure \ref{Fig:Kipp} shows the abundance profile of the 8904 M$_\odot$ star at the end of core Helium burning stage. The convective core extends until the mass coordinate of 1142 M$_\odot$ beyond which the effects of large intermediate convective zones become dominant as the profiles of various elements is flat until 5800 M$_\odot$, hinting at strong mixing in these zones. From 5800 to 7800 M$_\odot$, small intermediate convective are present which also promote mixing but to a lesser extent, shown by the negative gradients of the abundance profile of all elements in the right panel of Figure \ref{Fig:Kipp}. Finally, a large convective zones extends from a radius of 1600 to 15848 R$_\odot$ and mass from 8000 to 8904 M$_\odot$, produces a flat abundance profile and affects the surface abundances of the star. As a result of a near-fully convective structure, all chemical species such as $\mathrm{^{1}H}$, $\mathrm{^{3}He}$, $\mathrm{^{4}He}$, $\mathrm{^{12}C}$, $\mathrm{^{13}C}$ , $\mathrm{^{14}N}$, $\mathrm{^{16}O}$, $\mathrm{^{20}Ne}$ and $\mathrm{^{22}Ne}$, shown in the right panel of Figure \ref{Fig:Kipp} have their abundances strongly affected compared to their initial abundances. Globally in the zone above the convective core, hydrogen has been depleted (being transformed mainly into helium) and the other chemical species have been synthesised.

Although such massive stars are expected to generate energy via the CNO cycle already during core hydrogen burning, see \citet{Woods2017, Woods2021}, the amount of CNO material available is not sufficient for any significant $^{14}$N production during this evolutionary stage. The production of $^{14}$N instead occurs during the core helium burning stage. The first step in this process occurs in the $^{4}$He rich convective core where $^{12}$C is produced via the triple-$\alpha$ reaction. The $^{12}$C is then efficiently transported to adjacent hydrogen burning shell via convection. This begins the second step where the presence of $^{12}$C in the hydrogen burning shell allows it to capture hydrogen and produce $^{13}$N, which emits a positron and produces $^{13}$C. In the final step, $^{13}$C captures a proton. This produces $^{14}$N. The intermediate convective zones are present around the hydrogen burning shell and extend towards the surface, thereby allowing the efficient transport of $^{14}$N in the stellar envelope. The variation of the nitrogen abundance in mass fraction inside the star as a function of the lagrangian mass coordinate is denoted by the red line in the right panel of Figure \ref{Fig:Kipp} where the abundance of $^{14}$N is highest at the hydrogen burning shell that begins at the mass coordinate of 1142 M$_\odot$ and is the third most abundant element (after $^{1}$H and $^{4}$He) at the surface. The production of $^{16}$O is similarly related to the generation of $^{12}$C which for PopIII stars mainly occurs during the core helium burning stage as seen above. Once $^{12}$C is produced, it undergoes $\alpha$ capture to produce $^{16}O$ which is also transported to the surface as shown by purple line in the right panel of Figure \ref{Fig:Kipp}. In the envelope we see thus material processed by both H- and He-burning. The CNO cycle is responsible for building nitrogen. The He-burning is responsible for making carbon and oxygen. The final abundances shown in the right panel of \ref{Fig:Kipp}  comes out from a complex interplay between nuclear reactions and convection.

\subsection{Integrated abundances}\label{Sec:Abund}

\indent In Table \ref{table:11models} we show the mean mass fraction of various isotopes in the ejecta of individual star with different final mass, $\bar{\rm X}_i$ of the 11 model stars as discussed in \S \ref{Sec:Context}. Firstly we see a clear trend in the ejected abundances for hydrogen and helium that relates to the increasing mass of stars. For example for 10\% mass loss parametrisation the integrated mass fraction of $^{1}$H left at the end of core hydrogen burning decreases from 0.36 to 0.17. This is due to the near fully convective structure of such stars that allow the mixing of hydrogen and therefore fuse more hydrogen during their lifetimes. As a consequence, the mean mass fraction of $^{4}$He increases with increasing mass. The abundance trends for the CNO elements vary significantly for individual models. Taking model 6127 as an example since it provides the most significant contribution amongst the 11 stars. We find that within model 6127, $^{14}$N is evenly distributed in the interior and the mean mass fraction varies from 0.039 in the outer 10\% of total mass to 0.060 when considering all mass above the CO core. A similar trend is seen for $^{12}$C where the mean abundances vary from 0.001 for the 10\% case to 0.004 when considering everything above the core. The overall quantity of $^{12}$C for all three mass-loss parameterisation is the lowest of all CNO elements and is due to the conversion of $^{12}$C to $^{14}$N and $^{16}$O during the core helium burning stage. The amount of integrated $^{16}$O varies from 0.006 at 10\% to 0.017 for the mass above the CO core. Once again, this highlights that most of $^{16}$O is stored in the interior of the stars. Upon inspecting various integrated mass fractions in Table \ref{table:11models}, a general trend concerning the increase or decrease in abundances as a function of mass can be observed, with the exception of models such as 1923. 
This is likely due to the accretion history of such models where periodic bursts in accretion after a quiescent phase occur well into core hydrogen burning phase. This injects fresh elements, for instance hydrogen in the star which leads to the envelope of such stars exhibhiting higher abundance of hydrogen. In summary, high mass stars such as model 6127 have the highest mass fraction of their chemical content in form of $^{4}$He and also have a significant mass fraction of $^{14}$N when compared to other CNO elements. $^{1}$H plays a much larger role in the overall mass fraction content of low mass models such as 491 and these models consist of larger mass fraction of $^{16}$O in comparison to $^{12}$C and $^{14}$N.

\begin{figure*}
   \centering
    \includegraphics[width=18cm]{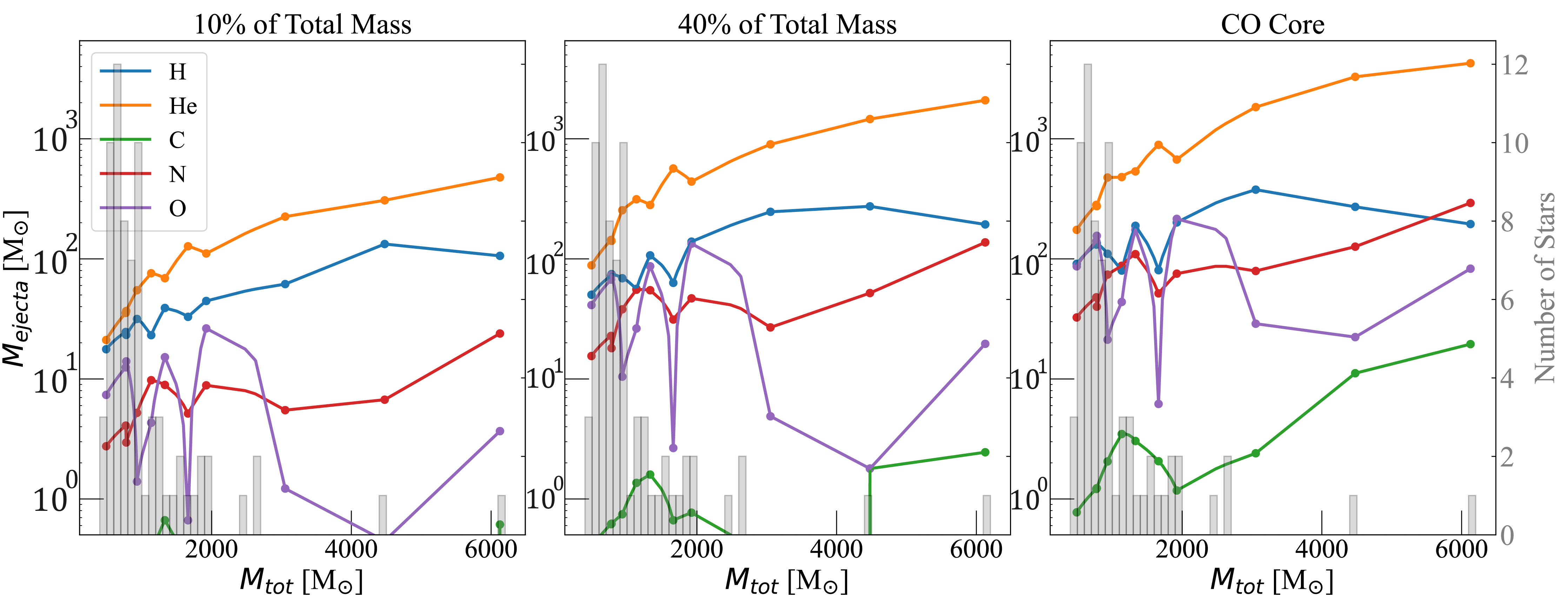}
      \caption{
      Mass ejected of various chemical species by each individual star in the halo from \citet{Regan_2020b}. The first y-axis with lines and dots shows the ejected masses in M$_\odot$ 
      for a given element for all the individual stars in the simulation of \citet{Regan_2020b}. The dots connect the interpolated values between the 11 computed models and each dot on lines correspond to the 11 models computed until the end of core Helium burning. The second y-axis represents the number of stars in the histogram only. The bins in the histogram cover a width of 100 M$_{\odot}$ starting from 40 M$_{\odot}$ and ending at 6200 M$_{\odot}$. The left panel shows the case when 10\% of the total mass is ejected. The middle and right panel, the cases when 40\%, and respectively the whole mass above the CO core is ejected.}
         \label{Fig:M_ejecta}
   \end{figure*}

To determine the mass of the different elements ejected by other initial mass stars (not included among the 11 cases that have been computed here), formed in the simulation of \citet{Regan_2020b}, we perform linear interpolation for associating a CO core mass and an ejected mass of an isotope to the remaining 90 stars for which we did not compute a detailed evolutionary sequence. In Fig. \ref{Fig:M_ejecta},
we show how the ejected masses of different elements vary as a function of the mass of the different stellar models.

Additionally, a histogram representing the mass distribution of all stars from the cosmological simulation is over-plotted at the corresponding mass. A general trend concerning all elements is visible in the three panels; the total amount of any given element ejected (in M$_\odot$) increases as a larger and larger portion of star is considered for mass-loss. This is a quite obvious fact but it also shows that all elements are well mixed in the interior and envelope of all stars. As seen in all three panels of Figure \ref{Fig:M_ejecta}, the integrated ejected mass of $^4$He (orange line) increases as the mass of stars in the halo increases. This is due to the fact that as the mass of a star increases, larger portions of the convective core are able to sustain temperatures high enough for $^1$H to fuse into $^4$He, thereby depleting the star of $^1$H and producing an excess of $^4$He. The case of ejecting the entire mass above the CO core can be examined in right panel of Figure \ref{Fig:M_ejecta}, the total helium (orange line) produced by a single 6200 M$_{\odot}$ bin is an order of magnitude larger than any single 600-700 M$_{\odot}$ star, or is equal to all 12 stars in the 600-700 M$_{\odot}$ mass bin. \\
\indent The trend for ejected mass of $^1$H (shown as blue line in all panels of Figure \ref{Fig:M_ejecta}) is qualitatively similar to $^4$He but has a more complex profile for masses ranging from 1000-6000 M$_\odot$. The integrated ejecta mass of $^1$H increases for the left panel of Figure \ref{Fig:M_ejecta} but in case of the middle and right panels, the value peaks at around 3000 M$_\odot$ and then decreases. This is due to a more efficient core hydrogen burning shell that produces more $^4$He in stars with larger masses above 3000 M$_\odot$. \\
\indent The integrated ejected mass of $^{14}$N (shown as red lines in Figure \ref{Fig:M_ejecta}) also shows an increasing trend similar to $^4$He with the most massive stars producing the majority of $^{14}$N. However in the case of integrated ejecta for $^{16}$O (purple line), the highest amount is produced by stars of 2000 M$_\odot$ and decreases as mass is further increased. This drop is more significant in 10\% and 40\% cases and shows that most of $^{16}$O is locked in the deep interior of the star and is less mixed when compared to other elements. The integrated ejecta for $^{12}$C (green line) also appears to be locked in the deep interior, as seen when the case of CO core (right panel of Figure \ref{Fig:M_ejecta}) is considered. Here $^{12}$C integrated ejecta mass follows a trend consistent with other chemical species but is similar to $^{16}$O in 10\% and 40\% cases. These trends are due mainly to the fact that nitrogen results from nuclear processes occurring in the H-burning shell, while carbon and oxygen comes from the deeper He-burning regions and also from the fact that more massive a star is, in general larger mixed zones appear.

\begin{table*}[h]
\caption{Table depicting the abundance ratios of various chemical species of the entire halo. The first column indicates the assumption for the mass lost. The columns 3 to 9 indicate the mean abundance (in mass fraction) of various isotopes in the ejected material by all the stars
in the halo simulation of \citet{Regan_2020b} with masses above 490 M$_\odot$. The quantities in these columns correspond to $X_{ip}$ defined by Eq.$~$\ref{eq:4}. The next five columns show the valuos of log$_{10}(N/O)$, log$_{10}(C/O)$, log$_{10}(O/H) + 12$, log$_{10}$($^{12}$C/$^{13}$C) and log$_{10}$($^{22}$Ne/O) respectively, in the material composed from the sum of the ejecta of all the stars. The ratios are in number fractions.}
\centering
\resizebox{\textwidth}{!}{%
\begin{tabular}{|c|c|c|c|c|c|c|c|c|c|c|c|c|c|}
\hline
Fraction & $M_{\text{{ejecta}}} [M_{\odot}]$ & $X_{\rm H1,p}$ & $X_{\rm He4,p}$ & $X_{\rm C12,p}$ & $X_{\rm N14,p}$ & $X_{\rm O16,p}$ & $X_{\rm C13,p}$ & $X_{\rm Ne22,p}$ & log$_{10}$(N/O) & log$_{10}$(C/O) & log$_{10}$(O/H)+12 & log$_{10}$($^{12}$C/$^{13}$C) & log$_{10}$($^{22}$Ne/O)\\
\hline
10\% of $M_{\text{tot}}$ & 8900.7 & 0.35 & 0.55 & 0.003 & 0.04 & 0.06 & 0.0004 & $5.7\times10^{-5}$ & -0.17 & -1.30 & 11.23 & 0.875 & -3.02\\
\hline
40\% of $M_{\text{tot}}$ & 32602.8 & 0.22 & 0.63 & 0.009 & 0.065 & 0.08 & 0.0005 & $7.7\times10^{-5}$ & -0.09 & -0.95 & 11.56 & 1.25 & -3.02\\
\hline
Above CO core & 65866.5 & 0.20 & 0.65 & 0.012 & 0.07 & 0.07 & 0.0004 & $8.8\times10^{-5}$ & 0.00 & -0.76 & 11.54 & 1.47 & -3.13\\
\hline
\end{tabular}
}
\label{table:HaloA}
\end{table*}
\begin{figure*}
   \centering
    \includegraphics[width=18cm]{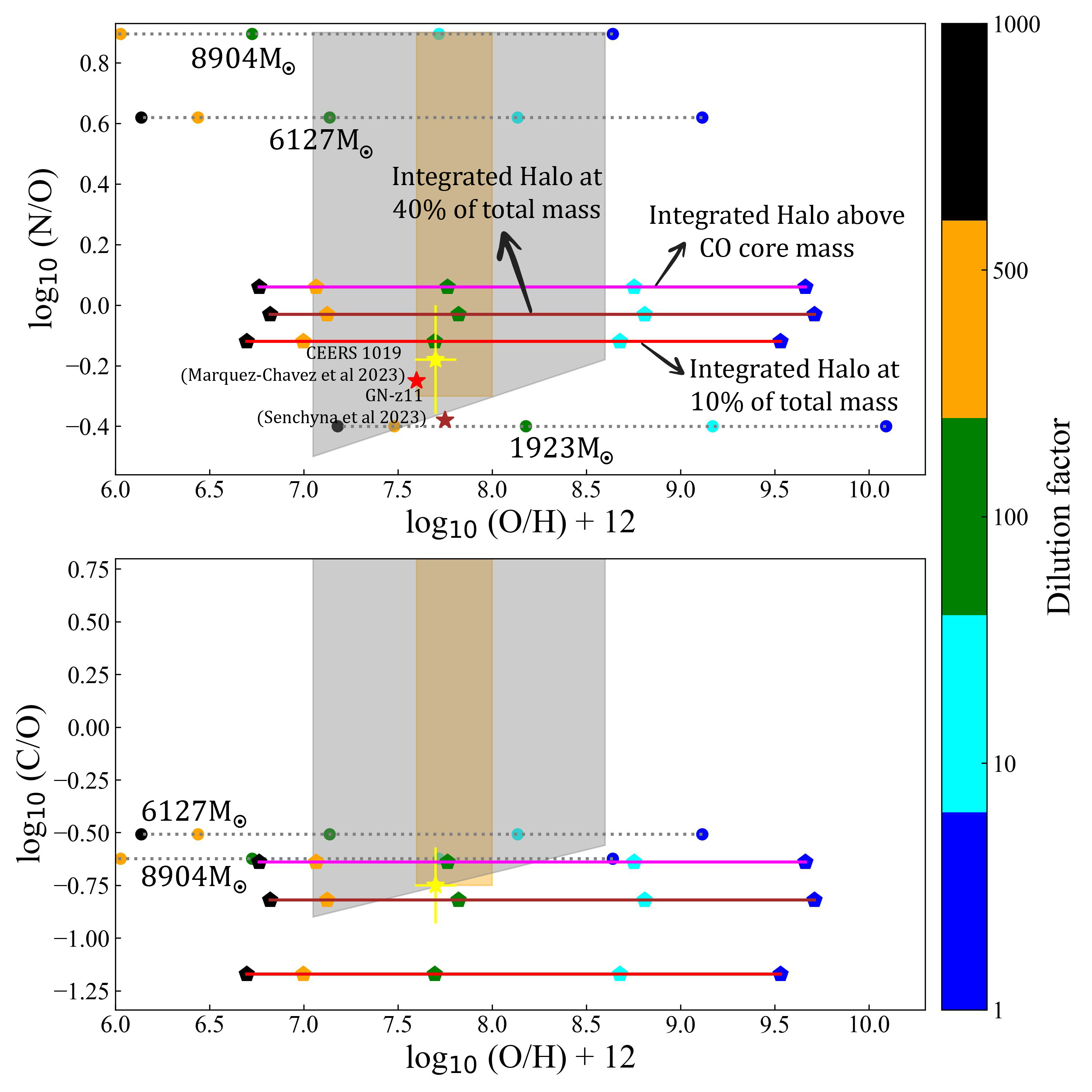}
      \caption{Abundance ratios (in number fractions) in the interstellar medium of the halo as a function of O/H (expressed as log$_{10}$(O/H)+12). The dilution factor, $f$, is given by the colourbar. 
      Dotted lines correspond to the abundance in the ejecta of an individual stellar model. The mass is indicated for each of them. The result
      for dilution factors equal to 1, 10, 100 and 1000 are shown by coloured circles. 
        The solid lines show the abundances when the ejecta of all the stars in the halo are mixed and mixed also with varying amounts of pristine material: magenta line is for the case when the entire mass above the CO core is assumed to be ejected by each star, brown and red lines for the cases when 40\%, respectively 10\% of the total mass is ejected. The coloured pentagons on the solid lines show the results for different dilution factors. The grey trapezoid and the orange rectangle represent the conservative and fiducial results for the abundances of GN-z11 by \citet{Cameron2023}. The yellow star with error bars correspond to the results of CEERS 1019 as shown in \citet{Marques2023}.  
      \textit{Top Panel}: The top panel shows the N/O abundances as a function of the O/H abundances both in log base 10 scale. The red and brown stars in the top panel are the photoionisation modelling results for GN-z11 by \citet{Senchyna2023}. {\it Bottom Panel:} C/O versus O/H ratios represented in the log base 10 scale. }
         \label{Fig:Results}
   \end{figure*}

Finally, to explore whether such a (progenitor) galaxy could produce the high abundances observed in candidates such as GN-z11 and CEERS 1019, we sum up the contributions of all the individual stars composing the population  to one isotopes and obtain the values in Table \ref{table:HaloA}. The first striking feature is the extremely high log$_{10}$ N/O mass fraction ratio (-0.17 for 10\% of total mass and 0.00 for the mass above CO core) and along with high O/H ratios (11.23 for 10\% of total mass and 11.54 for the mass above CO core). On the other hand, the C/O ratio is quite low. This is due to the strongly depleted $^{12}$C in all models (-1.30 for 10\% of total mass and -0.76 for the mass above CO core). Driving the depletion of $^{12}$C relative to $^{14}$N and $^{16}$O is the fusion of $^{12}$C to produce the other two elements during the core helium burning stage and is similar to the trend observed in massive stars, as seen in the works of \citet{Murphy2021a}.   
\subsection{Impact of dilution factor}\label{Sec:Abund}
The abundance ratios obtained from individual models and the entire galaxy represent the amount that will be ejected by the respective object(s) and take no account of pre-existing enrichment. However, in order to account for pre-existing enrichment of the ISM, a dilution factor ($f$ = $M_{\mathrm{ISM}}/M_{\mathrm{EJ}}$
) is included in further calculations. Due to the complexity in determining the precise value of the dilution factor, we assume values of 1, 10, 100, 500 and 1000. The results are then compared to the observational data for GN-z11 and CEERS 2019 and are shown in Figure \ref{Fig:Results}. In Figure \ref{Fig:Results} we show a selection of individual models as well as the results from when all of the stars are included. The model with a final mass of 6127 M${_\odot}$ reproduces the N/O, C/O and O/H values very well over a wide range of dilution factor ($f = 10 \rightarrow 100$). The values for the composition of the ejecta coming from all the stars using three different cutoff values can also reproduce the N/O number fractions of -0.13, -0.04 and 0.05 for 10\%, 40\% and above the CO core respectively - entirely consistent with observations. However, the C/O number fraction (bottom panel) of -0.64 is the closest of the observed ratios in the two considered high redshift galaxies when the entire mass above the CO core is assumed to be ejected. The O/H number fractions of 7.70, 7.80 and 8.75 for 10\%, 40\% of total mass and all mass above the CO core for the entire galaxy come close to the observations if a dilution factor of 100 is considered. A dilution factor of 100 is exactly what we expect for a 10\% ejection fraction combined with a star formation rate of 0.1. This is compatible with a situation in where the masses of stars represents only 10\% of the total mass (stars and gas) of the clusters, and where 10\% of the stellar mass is ejected.

\subsection{Massive PopIII stars as efficient Nitrogen enrichers and the progenitors of high-z N-loud Quasars}
The evolution of accreting massive Pop III stars demonstrates profound nitrogen enrichment of the outer envelope, as depicted in Figure \ref{Fig:Kipp}. Such enrichment is principally driven during the core helium burning phase, where the presence of $^{12}$C advances the production of $^{14}$N. However, these stars tend to nearly exhaust their \(^{12}{\rm C}\) reservoirs in this enrichment process. Delving into the integrated abundances across an entire (sub-)galaxy population, there's an evident correlation: as stellar mass increases, the remaining fraction of $^{1}$H decreases while $^{4}$He exhibits the opposite trend. Specifically, in the first galaxy simulations of R20b, the stellar mass range leads to remarkably high N/O and O/H ratios, with a diminished C/O ratio, attributed to the pronounced carbon depletion across models. Introducing the concept of dilution to assess the impact on the interstellar medium, the 6127 M${_\odot}$ model aligns very well with observed ratios such as N/O, C/O, and O/H over a dilution factor spectrum of 10-100. Moreover, considering realistic mass cutoffs and a dilution factor of 100, the integrated values from all of the massive stars agrees very closely with observational benchmarks as shown in Figure \ref{Fig:Results}. What this tells us is that a sub-population of very massive PopIII stars can reproduce the elemental abundances as seen by JWST and in particular explains very well the unusually high nitrogen abundances.

\section{Discussion}\label{Sec:Discussion}

\subsection{Mass ranges}\label{Sec:Range}
The chemical signatures exhibited by the massive PopIII stars are strongly dependent on their mass during the various mass-loss events and at the time of collapse. Figure \ref{Fig:Results} shows both the impact from the individual stars from the simulations of \citet{Regan_2020b} and those from all of the stars. An individual star with mass less than 1923 M$_{\odot}$ results in a log$_{10}$(N/O) ratio of -0.37, which is on the lower end of observed values. The log$_{10}$(C/O) ratio on the other hand misses the observed values entirely. The next mass in the simulation (not shown in Figure \ref{Fig:Results}) that produces a log$_{10}$(N/O) ratio of 0.1 and log$_{10}$(C/O) ratio of -0.7 is 2479 M$_{\odot}$.
There also exists an upper mass limit for haloA beyond 8904 M$_{\odot}$ where the log$_{10}$(N/O) ratio exceeds 0.85 and log$_{10}$(C/O) reduces below -0.6, thereby missing the observational window for GN-z11 and CEERS 1019. Such a conclusion is strictly time dependent as continual pollution of ISM by the next generations of stars would change the log$_{10}$(N/O) value. Interestingly supermassive PopIII stars with masses above 100,000 M$_{\odot}$ cannot produce the observed values as they experience a premature collapse during core hydrogen burning \citep{Lionel2018}. This evolutionary stage, prior to helium burning, is devoid of any substantial production of CNO elements which only occur during the later core helium burning stage. This result demonstrated that there exists both an upper and lower mass limit where massive PopIII are capable of producing the observed signatures. We note that this does imply that such massive stars do not exist - simply that stars in the mass range of approximately $10^3 - 10^4$ are required to re-produce the abundances signatures seen by JWST.

\subsection{Predicted abundances}\label{Sec:Predict}

Here we attempt to predict abundances of additional elements that have not yet been observed so as to make the model falsifiable. The goal here is to provide some additional values to better constrain future observations of objects such as GN-z11 and CEERS 1019. While our study primarily focuses on a specific stellar mass range, it represents only a fraction of the total stellar mass. However, the significance of our work lies in demonstrating the capability of these supermassive stars (SMSs) to both produce substantial amounts of Nitrogen and seed massive black holes. Additionally, various works have shown that  Late (z< 7) Pop III star formation might also have occurred in pristine regions due to in-homogeneous metal enrichment of the first galaxies \citep{Tornatore2007, Salvatera2011, Vanzella2020}. With this in mind, we find the integrated ejected mass in our simulated galaxy to be primarily enriched in helium with 0.55 mass fraction for 10\% of total mass and 0.65 for all mass ejected above CO core. This mass fraction allows us to predict the undiluted log$_{10}$(He/H) for 10\% and above CO core ratios of all stars for entire halo to be -0.40 and -0.09. Once the helium and hydrogen rich ejecta is exposed to the ISM, we can further predict the diluted log$_{10}$(He/H) ratios. For an equal contribution from the total stellar ejecta of all stars and the ISM, f = 1, we obtain log$_{10}$(He/H) = -0.62. For stronger dilutions of f = 10, 100 and 1000, the ratio decreases to -0.99, -1.07 and -1.08 respectively. Given that observing He/H ratios is an important observable signature for massive stars \citep{Schaerer2003}, these predicted ratios are useful for future JWST observations.

The log$_{10}$($^{12}$C/$^{13}$C) is found to be 0.87 for the 10\% case and 1.47 for the CO core case. To provide a comparison, 
the solar composition material shows a ratio 
$^{12}$C/$^{13}$C  around 60 \citep[see Table 2 in][and references therein]{Ekstrom2012}. Once these isotopes reach an equilibrium state (i.e. when the process has reached a stationary regime and the abundances of the CNO elements no longer vary) in material that has been processed by CNO burning reactions, the ratio
decreases to value around 4. The $^{12}$C/$^{13}$C ratios from the integrated population when considering 10\% and all mass above CO cases at the end of core helium burning are found to be 8 and 25 respectively, which clearly show the sign of CNO burning. It is also important to note that the absolute values of both $^{12}$C and $^{13}$C are boosted by the mixing between the helium and hydrogen burning regions. Fast rotating massive stars would also produce $^{12}$C/$^{13}$C ratios around similar values as the very massive stars considered here \citep{Chiappini2008}.

The right hand panel of Figure \ref{Fig:Kipp} shows that Neon abundance of such stars is strongly dominated by the $^{22}$Ne isotope, which is formed via the fusion of $^{14}$N. However, such stars also have a high $^{16}$O abundance which reduces the overall log$_{10}$($^{22}$Ne/O) ratio to -3.02 for the 10\% case and -3.13 for the CO core case. Comparing these values to the log$_{10}$(Ne/O) ratio of -0.63 obtained by \citet{Marques2023}, we find that PopIII stars in a halo such as ours cannot produce the observed Ne/O ratios, at least for galaxies such as CEERS 1019. To further explore the possibility of such massive stars producing the required Ne/O ratios, we invoke a scenario where the very and extremely massive stars form in an already enriched halo by previous generations of stars. We compute one model with a final mass of 3200 M$_\odot$ at Z = 10$^{-6}$ until the end of core helium burning and calculate the mean abundances in mass fractions for various elements in its ejecta, assuming all the mass above the CO core is lost at the end of its evolution. Upon ejection of the chemical species by this model, the ejecta is mixed with the non zero metallicity ISM with a dilution factor ranging from 1 to 1000. The results for a dilution factor 100 show a number fraction ratio of log$_{10}$(Ne/O) = -0.58. Additional number fractions of CNO are; log$_{10}$(N/O) = -0.52, log$_{10}$(C/O) = -1.29 and log$_{10}$(O/H)+12 = 8.03. The N/O and C/O ratios for this single model is lower than the observed ratios for GN-z11 and CEERS 1019 but such ratios will likely be different if a population of such stars with different masses are considered in a given halo (as we did for the Pop III case). At least, the numerical example shown here indicates that there is good hope that starting from a non-zero metallicity would still allow strong enrichment in CNO elements compatible with high redshift star forming zones, while also compatible with the high Ne/O ratio.

\subsection{Impact of a population of 20 M$_{\odot}$ stars} \label{Sec:Massive}
Due to the constraints in resolution of the hydrodynamic simulations of \citet{Regan_2020b}, stars with masses less than 40 M$_\odot$ are likely to be unresolved but should exist in real proto-galaxies of this mass. We can nonetheless estimate their impact on the N/O, C/O and O/H abundances by assuming 100 PopIII stars with 20 M$_{\odot}$ are born in addition within the halo. This is likely to be a very conservative estimate. Although the outcome of core collapse of 20 M$_\odot$ is realistically unknown, we assume these 20 M$_{\odot}$ models without rotation to undergo classical core-collapse which we expect will eject a huge portion of the star beyond the CO core, see works by \citet{Murphy2021}. 

The ejecta from an individual 20 M$_{\odot}$ will contain 9.6 M$_{\odot}$ of $^{1}$H, 6.2 M$_{\odot}$ of $^{4}$He, 1.4 M$_{\odot}$ of $^{16}$O, 0.6 M$_{\odot}$ of $^{12}$C and merely 1.2$\times 10^{-7}$ M$_{\odot}$ of $^{13}$C. 
Using these values, we can estimate the impact of 100 of 20 M$_{\odot}$ stars on the mass fraction values of log (N/O), log (C/O) and log (O/H) found in table \ref{table:HaloA}. For the case of 10\% of the total mass ejected from all stars without accounting for dilution ($\mathbf i$ = 0), we find a value of log (N/O) =  -0.27, log(C/O) = -0.87 and log(O/H) + 12 = 11.21. These values when compared to row 1, column 10, 11 and 12 of Table \ref{table:HaloA} show a decrease in log (N/O) mass fraction  values (from -0.17 to -0.27), an increase in log(C/O) values (from -1.30 to -0.87) and a near identical log(O/H) + 12 values (from 11.23 to 11.21). However, the impact of adding these 100 20 PopIII stars would be smaller for the cases 40\% and `above the CO core'.
Conclusively, these values are not significantly affected by 100 20 M$_{\odot}$ stars however if the halo were to contain an excess of a 10$^4$ stars, the impact will be enhanced. However, addition of 20 M$_\odot$ PopIII stars would improve our agreement for the log (C/O) and log (Ne/O) ratio in the bottom panel of Figure \ref{Fig:Results}.

\subsection{Comparison with previous studies}\label{Sec:Comparison}

The presence of strong NIII] $\lambda$ 1750 and CIII] $\lambda\lambda$1909 lines observed in GN-z11 by \citet{Bunker2023} present unusually high N/O ratios at a redshift of z = 10.6. Using the emission line fluxes from \citet{Bunker2023}, \citet{Cameron2023} derived the abundance ratios quantitatively and found log (N/O) > -0.25, log (C/O) > -0.78 and log (O/H) $\approx$ 7.82 for their fiducial model and log (N/O) > -0.49, log (C/O) > -0.95 and log (O/H) $\leq$ 8.60 for their conservative models as denoted by orange and grey box in Figure \ref{Fig:Results}. 

Using the runaway collision scenario proposed by \citet{Gieles2018}, \citet{Charbonnel2023} were the first to explore the N/O, C/O and Ne/O of GN-z11. Using a SMS model with a characteristic mass of 10$^4$ M$_{\odot}$ and metallicities ([Fe/H]) ranging from -1.76 to -0.78. The mechanism of transport of chemical species in their work is the presence of a fully convective SMS that grows via collisions with smaller objects throughout its lifetime. Additionally, they predict the Ne/O in GN-z11 to be higher than other high-z objects when compared to normal galaxies with similar metallicity. In contrast we invoke a more conservative massive PopIII formation scenario where the transition to the later stages of evolution, specifically post H-burning allows the structure to become convective and transport the elements to the surface more efficiently. Our single star models such as 6127 produce a higher N/O and C/O ratio whereas the O/H ratios match for a dilution factor between 10-100. Our integrated abundances for our total population of massive stars provide a closer match to the observed values of N/O and C/O, whereas the O/H matches if the dilution factor is around 100. We also find that the $^{22}$Ne/$^{16}$O ratio for the population to be around -3.0 which is much lower than the values of GN-z11 predicted in their work. 

\citet{Nagele2023} compute metal-enriched SMS models with rotation, stellar winds and SMS explosions to explain the high nitrogen pollution in GN-z11. The four models in their study were generated at ZAMS with initial masses of 10$^3$, 10$^4$, 5$\times$10$^4$ and 10$^5$. Each model was evolved either until the late carbon burning phase or until the GR instability was reached. The total $^{12}$C, $^{14}$N and $^{16}$O values were 9.91, 17 and 27 M$_{\odot}$ respectively for the 10$^3$ M$_{\odot}$ model. In comparison, $^{12}$C, $^{14}$N and $^{16}$O values for our 932 M$_{\odot}$ model are 0.741, 26.76 and 7.657 M$_{\odot}$ if all the mass above CO core is assumed to be ejected. The difference in the ejected masses is likely due to their inclusion of the effects of mass-loss, rotation and high metallicity in their models; our models are instead computed without mass-loss and rotation and have zero metallicity. Furthermore, \citet{Nagele2023} favour their high mass models when explaining the abundances of GN-z11 over a range of dilution factors whereas our models at 10$^5$ M$_{\odot}$ produce a log$_{10}$(N/O) = 2.2, log$_{10}$(C/O) = 0.45 and log$_{10}$(O/H)+12 varying from -2.3 to 0.8 over a range of dilution factors from 1 to 1000. This difference is possibly due to the difference in metallicity and structure of such objects. \\
\indent Finally, \citet{Kobayashi2023} advocate for a dual starburst model to explain the high Nitrogen abundances in GN-z11. By assuming that star formation in GN-z11 begins at z $\sim$ 16.7 and continued for 100 Myr before a quisient phase also lasting approximately 100 Myr \citet{Kobayashi2023} show that WR stars at [Fe/H] = 0, -1, -2 and -3 in the second star burst phase are potential sources of high (C,N)/O ratios. When considering their fiducial model following the dual starburst formation scenario and fixing the star formation age to match the log (O/H)+12 ratio , the log (N/O) ratio is found to be 0.246. They also conclude that very massive stars (VMS > 100M$_\odot$) with Z = 0 and 0.0002 as sources do not lead to a sufficient increase in N/O and pair instability supernovae (PISNe) lead to decrease in N/O ratio. They also show a single star burst model which requires stars with masses in excess of 1000 \msolar to explain the abundances - consistent with what we show here. \citet{Kobayashi2023} also concluded that the impact of adding
normal massive stars (typically of 20 M$_\odot$, as we did here) does not affect the overall N/O ratio of GN-z11.\\
\indent Another high-z JWST galaxy discovered by the CEERS survey team - CEERS 1019 - displays similar abundances to GN-z11. Investigating CEERS 1019 \citet{Marques2023} found strong emission of NIII and NIV lines and obtained log (N/O) = -0.13 $\pm$ 0.11, log (C/O) = -0.75 $\pm$ 0.11, log (O/H)+12 = 7.78 $\pm$ 0.18 and log (Ne/O) = -0.63 $\pm$ 0.07. Furthermore, \cite{Marques2023} invoke (super-massive) stars with masses in excess of 1000 \msolar to explain the unusual abundances (consistent also with similar anomalies observed in present day globular clusters \citep[e.g.][]{Denissenkov2014, Gieles2018}). 

\section{Conclusion}\label{Sec:Conclusion}

The research presented here has delved into the intricacies of massive PopIII stars and their pivotal roles in influencing the chemical constitution of specific galaxy environments. Through a meticulous analysis of an expansive dataset and detailed simulations, the study elucidates how the mass, evolution, and environmental interactions of massive PopIII stars are instrumental in deciphering the observed chemical signatures from early epochs of the Universe. Herein, we encapsulate the salient findings and their ramifications for forthcoming research.

\begin{itemize}
    \item \textbf{Mass ranges' importance}: The mass of the PopIII stars was found to be a determining factor in their resultant chemical signature. Specifically, stars less than $\sim$\(1900 \, \mathrm{M}_{\odot}\) produce a \(\log_{10}\) (N/O) ratio of -0.37, which is on the lower end of observed values. Stars beyond \(8900 \, M_{\odot}\) exceed the observed ratios for GN-z11 and CEERS 1019 and hence may be rare, at least in these galaxies. Only massive PopIII stars within the range of \(1900 \, \mathrm{\rm M}_{\odot}\) to \(8900 \, \mathrm{M}_{\odot}\) appear to align with the observed chemical signatures.

    \item \textbf{A population of very massive PopIII stars}: A sub-population of very massive PopIII stars produces an abundance pattern of nitrogen which is consistent with current observations by JWST of GN-z11 and CEERS 1019. The high nitrogen abundances are driven solely by the most massive stars among that population. Lower mass stars will dilute that abundance pattern as they do not produce sufficient Nitrogen to be consistent with observations. Hence, 
    it appears that some early galaxies must contain a minority population of very massive PopIII stars (i.e. stars with masses in excess of 1000 \msolarc) in order to be consistent with observations. 
    
    \item \textbf{Predicted halo abundance}: The mass of individual massive PopIII stars significantly influences the chemical outputs of entire halos. For instance, the integrated ejected mass due to all of the massive stars in our model resulted in enrichment in $^4$He with a mass fraction of 0.55 for 10\% of total mass and 0.65 for all mass ejected above the CO core. As a result, the integrated ejected mass of $^1$H varied from 0.35 to 0.20. Hence, our model predicts H
    helium rich environments in these embryonic galaxies. Additionally, low mass (M$_* \lesssim 2000$ \msolarc) PopIII stars are found to be the strong polluters in $^{16}$O whereas the ejecta of stars with masses above 2000 M$_\odot$ are enriched in $^{14}$N. Overall, after $^{4}$He and $^{1}$H, $^{14}$N dominates the ejecta of high mass PopIII stars. 

    \item \textbf{Matching abundances with GN-z11 and CEERS 1019}: If all massive PopIII stars eject a fraction of their total mass, we can closely reproduce the log (N/O), log (C/O) and log (O/H)+12 ratios over a range of dilution factor. Our massive PopIII models predict that N-loud quasars should be a key signature of a previous episode of massive PopIII star formation with the seed for the quasars possible originating from one of the PopIII stars responsible for the initial nitrogen enrichment. 

    \item \textbf{The origin of the high Ne/O ratio in CEERS 1019:} As mentioned above, our model produces very low Ne/O ratio (-3.02) compared to CEERS 1019 (-0.63). In the present models, neon is under the form of the $^{22}$Ne isotopes. It comes from the transformation of the $^{14}$N at the beginning of the core He-burning phase. The low value that we obtain indicates that this process alone is not efficient enough to produce Ne/O ratios compatible with the one observed in CEERS 1019. Additionally, we find that stars with mass around 3200 M$_\odot$ at 10$^{-6}$ Z can reproduce accurately the log$_{10}$(Ne/O) ratio of -0.58, but the N/O, C/O values are below the observed value. This concludes that it is difficult to reproduce all the observed ratios with just one generation of PopIII models; a more complex analysis of star formation history is needed.

    \item \textbf{Abundance predictions}: 
    It is important to emphasise that the current model predicts the abundances resulting from a first generation of extremely massive stars. As previously mentioned, these predictions are specifically applicable to a unique stage in the evolution of such systems. In this context, we also outline potential abundance signatures characteristic of these systems.
    Carbon ratios: The \(\log_{10}\) (\(^{12}\)C/\(^{13}\)C) ratio  for the 10\% case and CO core case was found to be 0.87 and 1.47 respectively, showing a strong signature of CNO cycle. The Ne/O ratios predicted in this study were found to be around -3.0 for the halo. These additional abundance ratios provide an estimated value for future observations of JWST. We encourage observers to seek out for these ratios in future observations.

    \item \textbf{Impact of 20 \( M_{\odot} \) stars}: Incorporating 100 stars of 20 \( M_{\odot}\) mass within the stellar population provides pivotal insights into the role of these less massive stars. Specifically, the contribution of these stars leads to \(\log\)$_{10}$ (N/O) and \(\log\)$_{10}$ (O/H)+12 values of -0.27 and 11.21 respectively, indicating their negligible effect on these ratios (i.e. PopIII stars of this mass cannot account for anomalously high N abundances). However, the addition of these stars into the halo increased the \(\log\)$_{10}$ (C/O) from -1.30 to -0.87 and would likely increase the Ne/O ratio, thereby providing a better fit to the observed values of GN-z11 and CEERS 1019.

\end{itemize}

As future studies continue to explore the possible scenarios for high red-shift candidates such as GN-z11 and CEERS 1019, our models robustly predict significant N enrichment from massive PopIII stars. If such stars exist, and they eject mass, it would be entirely unsurprising to find strong N-enrichment in the very early Universe. 
It is important to recognise that nitrogen enrichment can also result from fast-rotating massive stars with masses ranging between 8 and approximately 100 M$_\odot$ \citep{Meynet2006, Chiap2003, Chia2006, Chia2013}. Fast rotating models have been suggested to account for the value of about -1.5 (log(N/O)) observed in iron-poor halo stars. This value is more than an order of magnitude lower than what is displayed by high-redshift galaxies such as GN-z11 and CEERS 1019. However, it is worth noting that the chemical enrichment model for the halo of our Galaxy significantly differs from the one discussed here. In future research, it would be insightful to explore the influence of fast-rotating massive, very massive, and even supermassive stars within this context

\noindent The influx of observations by JWST via the JADES \citep{Maiolino_2023b, Maiolino_2023a}, CEERS \citep{Larson_2023} and UNCOVER \citep{Goulding_2023} surveys have revealed the existence of a population of massive black holes with masses ranging from as low as $4 \times 10^5$ \msolar up to $\lesssim 10^8$ \msolar at redshifts as high as z = 10.6. The masses of these black holes as well as their sub-Eddington accretion rates, in most cases, appears to strengthen the case for a heavy seed formation pathway with the seed being a massive PopIII or SMS the most likely candidate \citep[e.g.][]{Natarajan_2023}. Our investigation here adds 
extra weight to the heavy seed argument by showing that massive PopIII stars enrich the environment surrounding the (mini)-quasar with metal abundances consistent with observations. In light of the above, it is paramount for future investigations to further probe into the multifaceted nature of massive PopIII stars, either to affirm, challenge, or elaborate on the narratives and interpretations offered in this study.\\

\section*{Acknowledgements}

 D.N., S.E. and G.M. have received funding from the European Research Council (ERC) under the European Union's Horizon 2020 research and innovation programme (grant agreement No 833925, project STAREX). JR acknowledges support from the Royal Society and Science Foundation Ireland under grant number URF\textbackslash R1\textbackslash 191132. JR also acknowledges support from the Irish Research Council Laureate programme under grant number IRCLA/2022/1165. T.E.W.\ acknowledges support from the National Research Council of Canada's Plaskett Fellowship. E.F. and G.M. have received funding from the SNF project 200020-212124.

\bibliographystyle{aa}
\bibliography{biblio}

\appendix
\section{Impact of evolutionary stages on the integrated abundances}\label{Sec:Ap_stages}
To explore the impact of late stages of evolution on the abundances, we calculated the integrated abundances of Hydrogen, Carbon, Nitrogen and Oxygen for the 3053 M$_{\odot}$ model as shown in table \ref{tab1test}. We find that the integrated abundances do no vary significantly beyond the core Helium burning phase. This is due to the very short timescale of the subsequent phases which barely alters the total composition of the model.Therefore the 11 models computed until the end of core Helium burning provide sufficienlty accurate results. The core is defined as the mass coordinate in star below which $10^{-3}$ mass fraction of the previous chemical species is left. For instance, CO core has $10^{-3}$ mass fraction of helium (Y$_c$) is left in the core.
\begin{table}[h]
\caption{Integrated abundance of various elements inside a 3053M$_\odot$ model at zero metallicity computed until the end of core Si burning. Column 1 indicates various evolutionary stages. The integrated mass fraction of a given element is presented in columns 2, 3, 4 and 5.}
\label{tab1test}
\centering
\resizebox{\linewidth}{!}{
\begin{tabular}{l | c | c | c | c}
\hline
\multirow{2}{*}{\textbf{Evolutionary stage}} & \multicolumn{4}{c}{$\begin{aligned}
    \rm{\bar{X}_{i}}=\int_{M_{\rm cut}}^{M_{\rm tot}} X_{i}(M_{r}) dM_r/M_{\rm EJ}
\end{aligned}$} \\
\cline{2-5}

Z = 0.0 & $^{1}$H & $^{12}$C & $^{14}$N & $^{16}$O \\
\hline \hline
End He burning      &0.1570&0.00160&0.0380&0.0130\\ 
End C burning      &0.1570&0.0008& 0.0330&0.0120 \\
End Si burning      &0.1570&0.0008& 0.0330&0.0120 \\
\hline
\end{tabular}
}
\end{table}
\end{document}